\documentclass[twocolumn,showpacs,amsmath,amssymb]{revtex4}
\usepackage{graphicx}
\usepackage{color} 

\begin{document}

\title{Circular invasion of fluid into a quenched disordered media}
\author{Y. C. Lin, K. Yun, and T. M. Hong}
\affiliation{Department of Physics, National Tsing Hua University, Hsinchu 30043, Taiwan, Republic of China}
\date{\today}

\begin{abstract}
Bacterial colonies, granular fingers, and surfaces generated by point invasion all involve circular patterns. As the pattern grows, the size of interface increases. And the Kardar-Parisi-Zhang and Edwards-Wilkinson equations break down because they all fail to conserve the fluid volume for a moving boundary. If we plot the interface against the polar angle, the wriggling of the interface wrinkles progressively with time. It shows convincingly that the noise spectrum behind the growth of a circular pattern is drastically different from the white noise commonly adopted for a planar interface. This ruins the universal power law found in the latter system.
A new model was proposed for our revisit of the Hele-Shaw experiment using a circular invasion.
Analytic solutions are obtained in excellent agreement with experimental results.
\end{abstract}


\maketitle

The properties related to surface growth have attracted much attention. In
the past two decades, scientists reported scaling behavior in many
experiments, like paper wetting \cite{paper,paper2}, burn front \cite{flame}, and fluid invasion into disordered media \cite{scale1,scale4,scale5} etc..
Surprisingly the surface roughness of systems with different growth
mechanisms turns out to share the same scaling law. In 1989, Rubio, Edwards,
Dougherty, and Gollub found the scaling behavior to obey a power law with
exponent $\alpha $, $w\sim L^{\alpha }$ \cite{scale1}. Similar to the
critical phenomena at phase transitions, this exponent is used to
distinguish different universality classes and characterize their scaling
behavior \cite{scale8}. Among the various models for the growth mechanism,
the Kardar-Parisi-Zhang (KPZ) model is perhaps the most
famous \cite{scale7}. It includes a nonlinear term to account for the askew growth in the
diffusion equation of the Edwards-Wilkinson (EW) model \cite{ew}. Several
papers have tried to introduce more accurate terms to describe the depinning
dynamics of a driven interface in an isotropic disordered media, such as the
quenched EW \cite{ew}, quenched KPZ \cite{scale7}, and quenched Herring-Mullins for a
tensionless interface \cite{qhm,scale9}. However, discrepancies are always there
when compared to the experiments (See Tab.I). So we decide to revisit the classic
Hele-Shaw experiment.

\begin{table}[tbp]
\caption{Comparison of the scaling exponent in the literature using a planar injection}
\label{tab1}%
\begin{ruledtabular}
\begin{tabular}{llll}
Experiment & $\alpha $ & Model & $\alpha $ \\ \hline
Rubio\cite{scale1} & 0.73 & KPZ\cite{scale7} & 0.5 \\ 
He\cite{scale4} & 0.65$\sim$0.91 & RFIM\cite{scale6} & 1 \\ 
Soriano\cite{scale5} & 0.94 & TOB\cite{scale2}/TBO\cite{scale3} & 0.7/0.6$\sim$0.9 \\
\end{tabular}
\end{ruledtabular}
\end{table}

\begin{figure}
\includegraphics[width=0.45\textwidth]{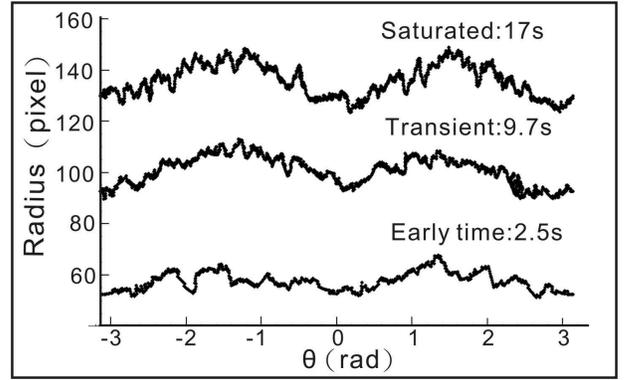}
\caption{\label{fig:demon}  Time-lapse interface profiles for invasion speed of $3.31\,\rm{cm^{2}/s}$. More local structures emerge with time.}
\end{figure}

We choose to inject the fluid
from a circular inlet in contrast to the standard planar one. Similar experiments do exist in 
the literature, but they often involve complications like the use of
non-Newtonian fluid or interface instabilities by prefilling the media
with another immiscible fluid \cite{radial,radial2}. Naively, we might expect the surface generated by a circular injection to eventually approach that of a planar one as the area of the invasion
liquid becomes large. We emphasize in this
Letter that this presumption is wrong mainly because the circumference of the circularly
injected fluid grows with time. This causes the constant introduction of 
new local structures as the surface expands, while old ones being pushed apart but, nevertheless, reserved (Fig.1).

Since the area of fluid is $\int R^2 d\theta/2$ for our circular invasion as opposed to
$\int h dx$ for the planar case, $R^{2}$ is a more natural quantity
to define the interface width than $R$. We first expand it in Fourier series: 
\begin{equation}
R^{2}\left( \theta ,t\right) =R_{0}^{2}\left( t\right)
+\sum_{l=1}^{\infty }\left[ a_{l}\left( t\right) \cos l\theta
+b_{l}\left( t\right) \sin l\theta \right]  \label{fourier}
\end{equation}
where $R_{0}$ is the mean radius of the whole interface. Then substitute it into the definition of interface width, 
\begin{equation}
w=\sqrt{\frac{1}{N}\sum_{n=1}^{N}\big{[}\frac{1}{\Delta \theta }\int_{( n-1) \Delta \theta }^{n\Delta \theta }R^{2}d\theta-( \frac{1}{\Delta \theta }\int_{( n-1) \Delta \theta}^{n\Delta \theta }Rd\theta ) ^{2}\big{]}}
\\
\nonumber \\
\end{equation}
where $\Delta \theta $ denotes the range of the angular average, and $N=%
2\pi /\Delta \theta $ is the number of divisions of the interface.
It is easy to show that the first term in the square root equals $%
R_{0}^{2}\left( t\right) $. The second term can be simplified if the
roughness is small enough ($a_{l},b_{l}\ll R_{0}^{2}$) to warrant the Taylor
expansion. In contrast to the random term in the KPZ model, we include the randomness by ensemble-averaging over the Fourier coefficients of Eq.(1) in the final expression for the interface width. Physically, this randomness comes from the distribution of spacing between neighboring beads along the advancing front. Overall, narrower spacings enjoy a stronger capillary force and thus a higher growth rate. Note that, in contrast to the white noise for a parallel injection which is uncorrelated in both sites and heights, the increasing size of the interface introduces chronologically high Fourier modes into the surface (Fig.1), which forces us to adopt a decreasing $l-$dependence for the noise,
$\langle a_{l}a_{l^{\prime}}\rangle=\langle b_{l}b_{l^{\prime }}\rangle=\zeta _{l}^{2}\delta_{l,l^{\prime}}$
and $\langle a_{l}\rangle=\langle b_{l}\rangle=0$ where the angular bracket denotes taking the statistical
average. 

\begin{figure}
\includegraphics[width=0.45\textwidth]{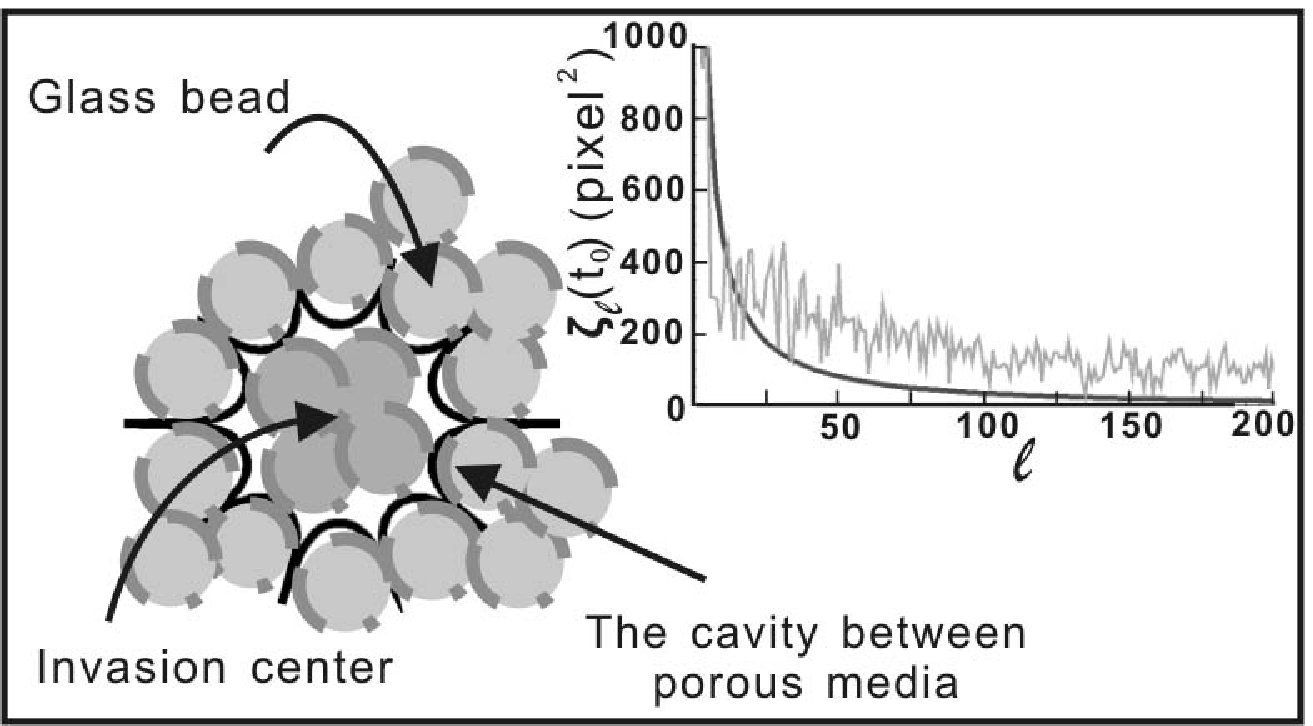}
\caption{\label{fig:initial} Schematic of the initial configuration of the invading fluid reconstructed from $\zeta_{l}(t_0)=1/(0.2+l+0.005l^2)$, obtained from the solid-line fit of the actual roughness spectrum $\zeta_{l}(t_0)$ with an invasion velocity of 5.1${\,\rm cm}^2/{\rm s}$ at transient times in the inset.}
\end{figure}

This allows us to drop most of the cross terms in the double
summation over $l$ and keep only the relevant terms. The final expression for
the width becomes
\begin{equation}
w\simeq \frac{1}{R_{_{0}}}\sqrt{\sum_{l=1}^{\infty }\left[ \frac{%
\zeta _{l}^{2}}{4}-\frac{\zeta _{l}^{2}\sin ^{2}(\frac{l\Delta \theta }{2})}{%
(l\Delta \theta )^{2}}\right] }
\end{equation}
Schematic plot of this result can be found in Fig.2. Note that when $%
\Delta \theta $ is small, this width always scales linearly with $\Delta \theta $ and
gives $\alpha =1$. In fact, this conclusion does not rely on the nature of noise spectrum. Without it, the equivalent expression of Eq.(3) becomes messier, but the lowest order term is still linear in $\Delta \theta$ when Taylor expanded. Since the next order term has a negative coefficient, it
becomes natural when we try to fit $\Delta \theta -c\left( \Delta \theta
\right) ^{2}$ by $\left( \Delta \theta \right) ^{\alpha }$ that an exponent
both less than 1 \cite{scale1,scale4,scale5} and time-dependent is obtained. So the eventual magnitude of this
exponent becomes arbitrary and varies with the range of $\Delta \theta$ 
we adopt for the power law fitting. This therefore casts doubts on the scaling arguments people conjure up to derive  $\alpha$ and conclude that it plays the same universal role as the exponents in the critical phenomena\cite{book}.

In order to study the dynamical behavior of Eq.(3), the actual dependence of $\zeta$ on $l$ is required. Note that our system size is growing in time. This is a moving boundary problem, and both the EW and KPZ equations fail to conserve the area. Instead of further complicating the existing models, we adopt the polar coordinates and modify the EW model to guarantee this conservation.  
\begin{eqnarray}
\frac{\partial R^{2}}{\partial t}=\nu \frac{\partial ^{2}R^{2}%
}{\partial \theta ^{2}}+\xi\left( t\right)\left[ R^{2}\left( t,\theta
\right) -R_{0}^{2}\left( t\right) \right]+\frac{V_{a}}{\pi }
\end{eqnarray}
where $V_a$ denotes the speed of fluid injection. Two phenomenological coefficients are introduced.
First, the relaxation coefficient $\nu$ is the same as in the EW and KPZ models, which derives from 
the surface tension and favors a smooth interface. The second term on the RHS is new. We ascribe it
to the inertia effect which tends to magnify the existing crests by allowing them to advance faster
in the next moment than the troughs. Therefore, we expect its coefficient $\xi$ to be proportional to
the average kinetic energy of the front, $\sigma \left( {dR_{0}}/{dt}%
\right) ^{2}\sim{\sigma V_{a}}/{t}$. It is interesting that Eq.(4), under the change of variable from $\theta$ to $x\equiv R\cdot \theta$, will generate a nonlinear term $(\partial R/\partial x)^2$ characteristic of the KPZ model.
Dimensional analysis allows us to pinpoint the dependence of these
two phenomenological coefficients on the microscopic parameters as: 
$\xi\left( t\right)=\xi^{'}/t=\kappa \sigma V_{a}/\eta r^{2}t$
and
$\nu={Tr}/{\eta }$
where $\sigma, \kappa, \eta, r, T$ are the fluid density, permeability, viscosity, bead size, and dynamic surface
tension. The reason why the contact angle did not come in is that we have limited the spacing between the two 
acrylic plates to be just a few beads to minimize its role.
Note that the source term in Eq.(4) vanishes when we plug Eq.(1) into Eq.(4). Applying the orthonormal relation, we end up with a simple linear
PDE for each of $a_l$ and $b_l$. They can be easily solved to determine the evolution of the average noise spectrum $\zeta_{l}$ as, 
\begin{eqnarray}
\zeta_{l}^{2}(t)=\zeta_{l}^{2}(t_0)e^{-2\nu l^{2}(t-t_{0})}(t/t_{0})^{2\xi^{'}}
\end{eqnarray}
Substituting the new noise form into Eq.(3), we obtain
\begin{eqnarray}
w=m\sqrt{\sum_{l=1}^{\infty}(\frac{1}{4}-\frac{\sin ^{2}(\frac{l\Delta\theta }{2})}{(l\Delta \theta )^{2}}) e^{-2\nu l^{2}(t-t_{0})} \zeta_{l}^{2}(t_0)}\\
m=\frac{N'sr^{2}}{2 R_{0}} \big{(}\frac{t}{t_{0}}\big{)}^{\xi^{'}}
\end{eqnarray}
where $N'$ is the fluid-covered area divided by glass bead cross-section and $\zeta_{l}(t_0)$ denotes the 
initial noise spectrum at $t_0$. Since there is no explicit random term in our PDE, this initial pattern cannot be taken as
a perfect circle, otherwise the subsequent evolution will not be able to produce any roughness. But, we also need to make
sure later that the final theory does not depend on the specific choice
of this initial time $t_0$ and its corresponding pattern $\zeta_{l}(t_0)$.
We considered several possible situations and generated the spectrum of the noise by analyzing the interface. 
The final form we adopt to describe the decay of noise with $l$ is $\zeta_{l}(t_0)=1/(s+l+s'l^{2})$ where the value of $s$ is found to fall between 0.1 and 1, and $s'$ is around 0.001.
This corresponds to an initial shape in Fig.2.


Our experimental setup includes a thin cell (0.15 cm deep) formed by a horizontally positioned  acrylic dish (radius=17 cm) and an acrylic top plate, a fluid injection system, and a data recorder. The syringe injects the fluid vertically into the disorder media from the center of the dish with speeds ranging from 2.19 to 2.50 ${\rm cm}^2/{\rm s}$. Glassbeads of diameter $500\,\mu \rm{m}$ are well tamped between the acrylic plates to make sure they will not move during the invasion. We use a high resolution (1 million pixels) CCD camera to monitor and record the interface, which is set vertically above the dish center. The camera monitors an area of $24\times24\,{\rm cm}^{2}$ plus the edge space, and the size of each pixel is about $450\,\mu{\rm m}^2$.
After the picture has been taken, a personal computer digitizes the video image and calculates the interface width.

Experimentally the weighting of the noise spectrum is found to decrease with the Fourier angular momentum. This is evidenced by Fig.1, which shows the interface wiggles progressively more severe with time. The high modes come in as more local structures sprout up at the newly gained surface. Lower modes, which correspond to more global structures, emerge earlier and thus have a longer history to evolve. According to the instability term in Eq.(4), fluctuations are amplified via feedbacks on themselves. Therefore, older structures tend to have larger amplitudes. The scale dependence of the width is shown in Fig.~\ref{fig:psresult}. 
The linear behavior of the width with a slope exactly equal to one in the asymptotic regime of small scales is consistent with we expected. Same features can be found in the bulk of non super-rough interface and many recent experiments \cite{scale5,scale10,scale11}.



\begin{figure}
\includegraphics[width=0.45\textwidth]{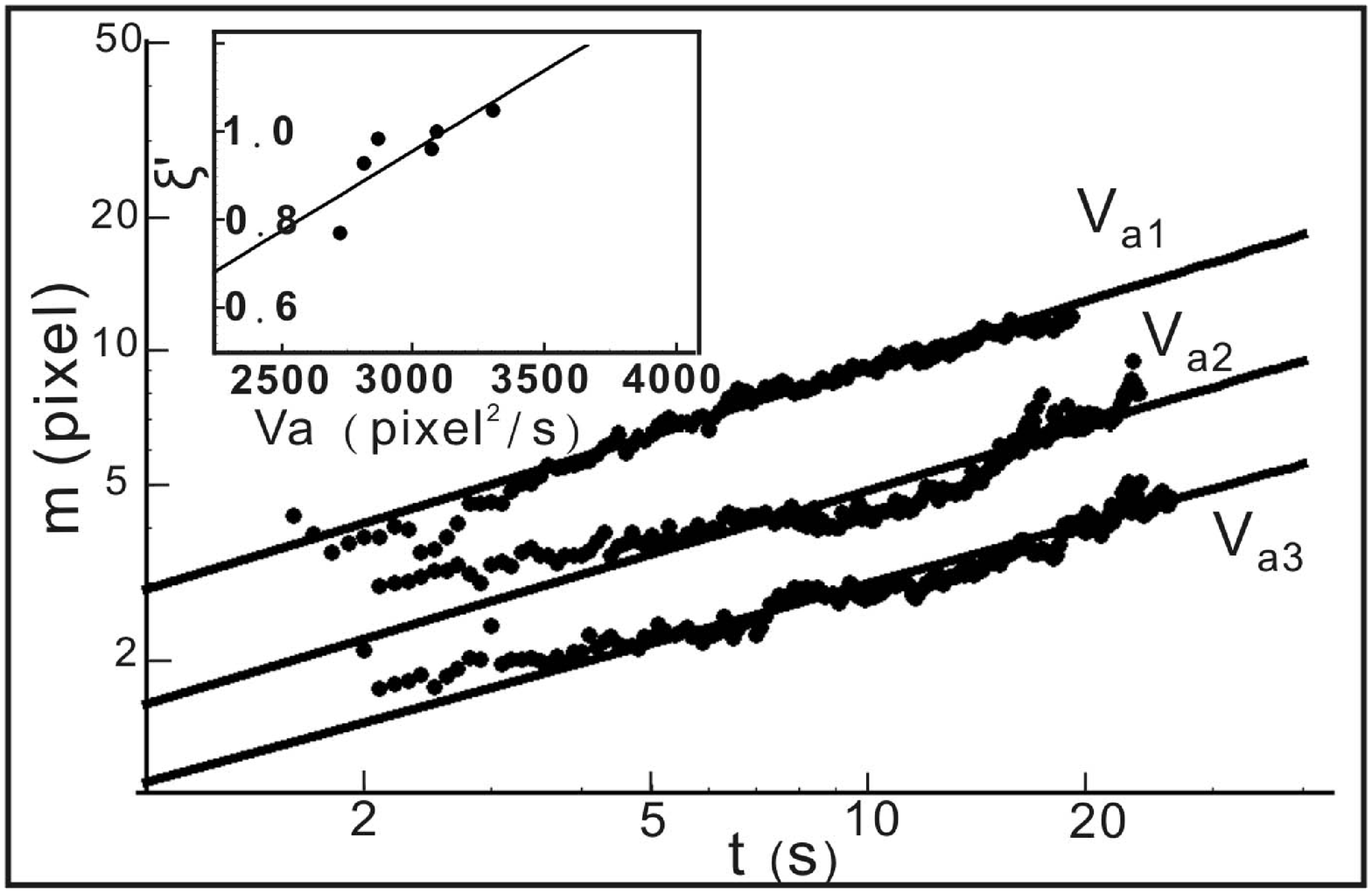}
\caption{\label{fig:instability}Logarithmic plot of the time dependence of $m$ defined in Eq.(9) for three invasion speeds, $V_{a1}=2.50$, $V_{a2}=2.40$, and $V_{a3}=2.19\,{\rm cm}^2/{\rm s}$. The solid line in the inset indeed passes through the origin when extrapolated.}
\end{figure}

\begin{figure}
\includegraphics[width=0.45\textwidth]{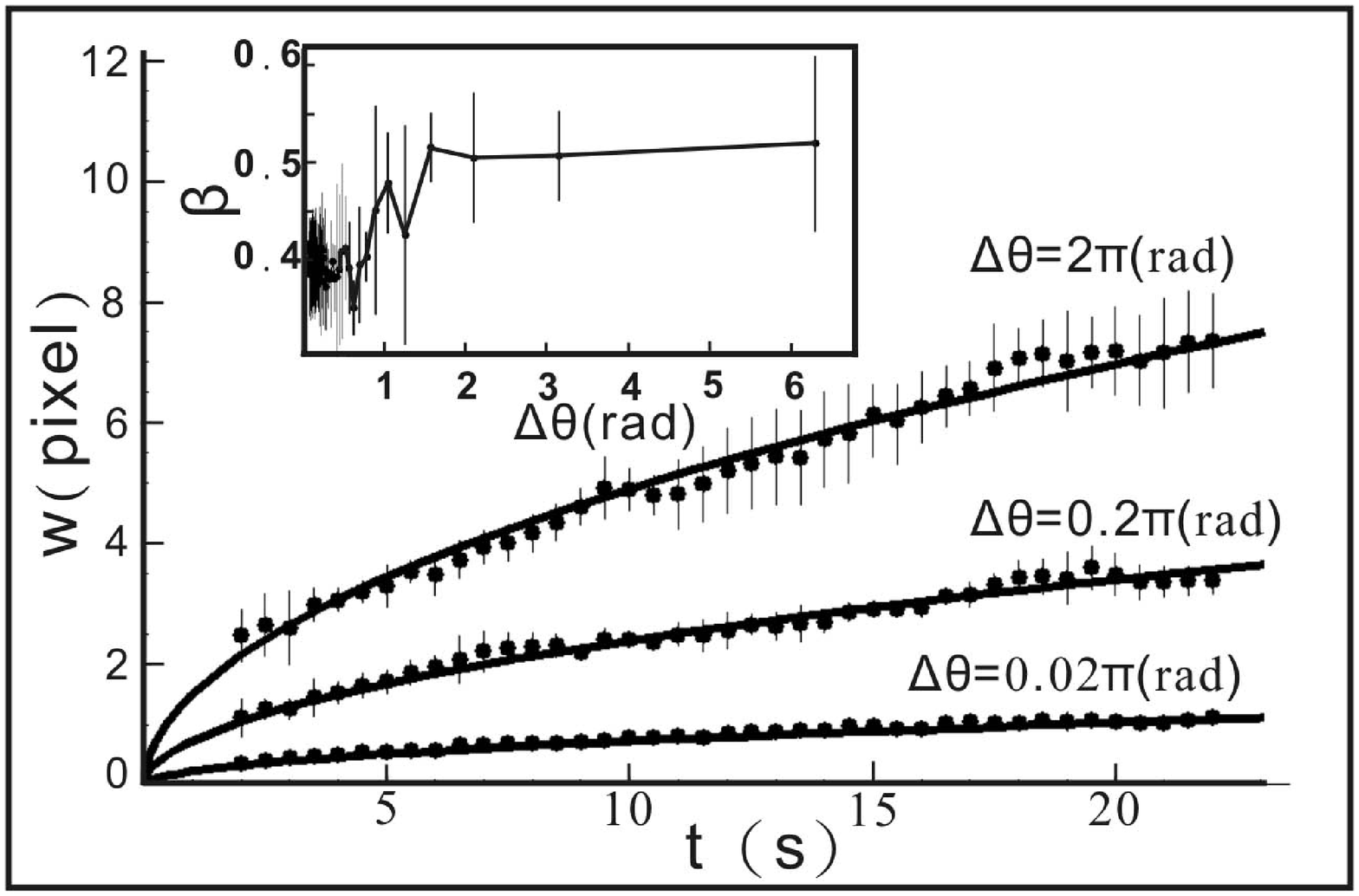}
\caption{\label{fig:pbresult}Interface width versus time for invasion velocity 3.15 $\rm{cm^{2}/s}$ under three different $\Delta\theta$. The forms of these three solid lines are determined by the model. The inset illustrate $\beta$ against $\Delta\theta$ for invasion velocity 3.15 $\rm{cm^{2}/s}$.}
\end{figure}

\begin{table}[tbp]
\caption{The values of the relevant parameters in the experiment at room temperature.}
\label{tab2}
\begin{ruledtabular}
\begin{tabular}{llll}
Parameter & Value (unit)\\ \hline
water-air surface tension ($T$)& $0.0727\,\rm{N/m}$\\ 
dynamic viscosity ($\eta$)& $1.002\times10^{-3}\,\rm{Pa\cdot s}$ \\ 
permeability ($\kappa$)& $\sim10^{-9}\,\rm{m^{2}}$\\
\end{tabular}
\end{ruledtabular}
\end{table}

 We use the parameters in Tab.II to estimate $\xi'$ and $\nu$ at about $1$ and $0.036$.
Experimental values put them at about $0.8\sim1.1$ (inset of Fig.4) and $10^{-2}\sim10^{-1}$ respectively. Other features are also verified experimentally, which include: (1) The instability coefficient $\xi'$ is linearly proportional to the invasion velocity (Fig.4), (2) no dependence on $V_{a}$ for the relaxation coefficient $\nu$. Moreover, Fig.5 finds $\Delta\theta$ dependence for $\beta$, as has been reported for a parallel injection\cite{scale5}.
By use of Eq.(6), we can obtain the initial noise spectrum $\zeta_{l}(t_0)$ by fitting the scale dependence of the interface width. Nicely, we deduce the same form at different times. Note that it is necessary to incorporate the possible symmetry of the initial configuration before being able to determine it directly from the spectrum in Fig.2. We duly checked that the interface width was not sensitive to the exact choice of this initial time $t_0$ and shape of interface as long as they belong to the early stage when the effect of randomness is still small. 
As illustrated above, both $\alpha$ and $\beta$ are smooth and varying functions of
the scale. On the other hand, $\xi'$ and $\nu$ are good parameters to describe the scaling and dynamic behavior of our circular interface.

\begin{figure}
\includegraphics[width=0.45\textwidth]{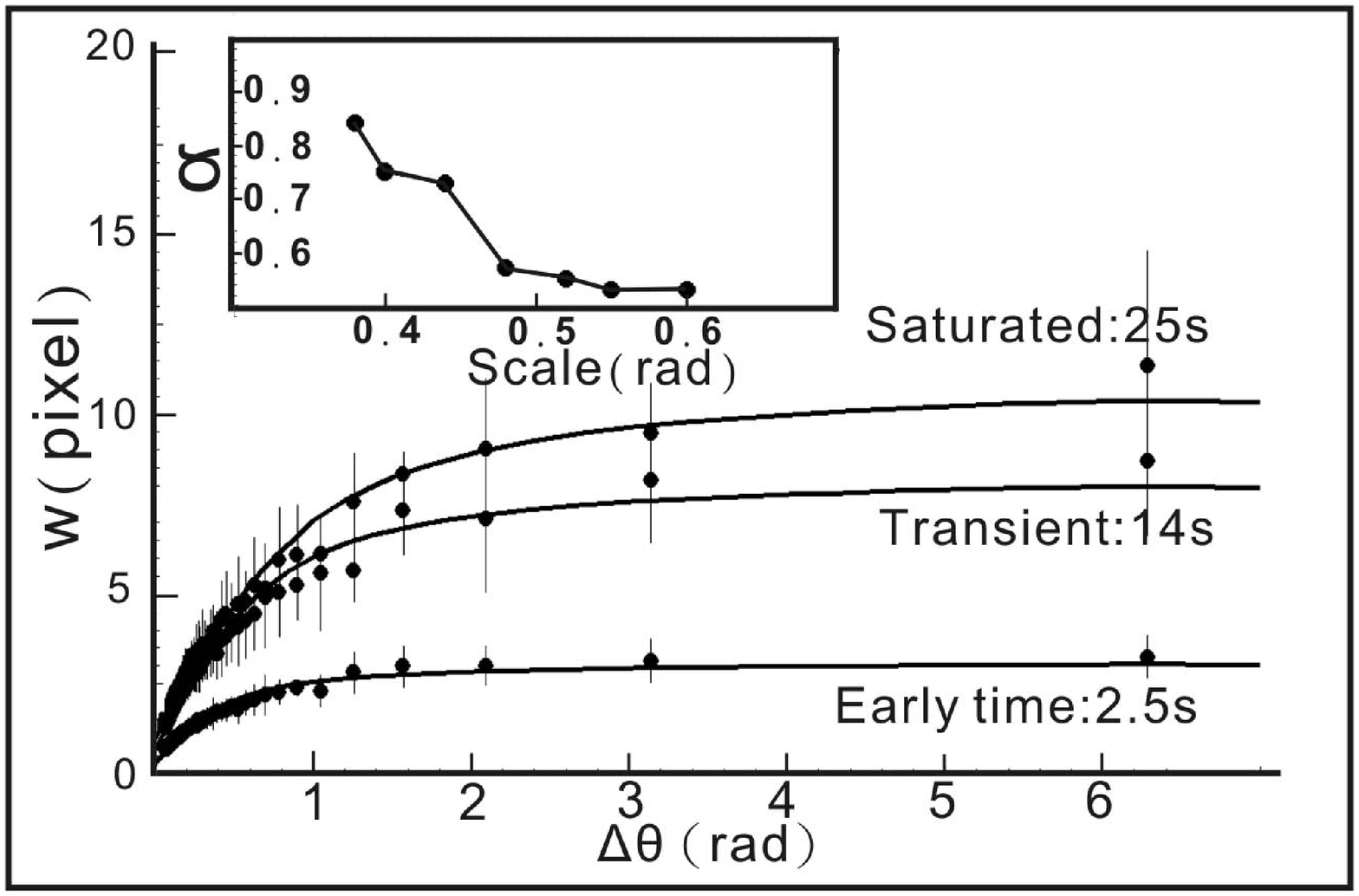}
\caption{\label{fig:psresult}Interface width versus $\Delta\theta$ at three different stages labelled by early time, transient, and saturation. The solid line comes from our formula. Invasion velocity is 5.1$\,{\rm cm}^{2}/{\rm s}$. The inset demonstrates that the actual value of $\alpha$ varies with the range of $\Delta\theta$ we use to do the power law fitting. }
\end{figure}

\begin{figure}
\includegraphics[width=0.45\textwidth]{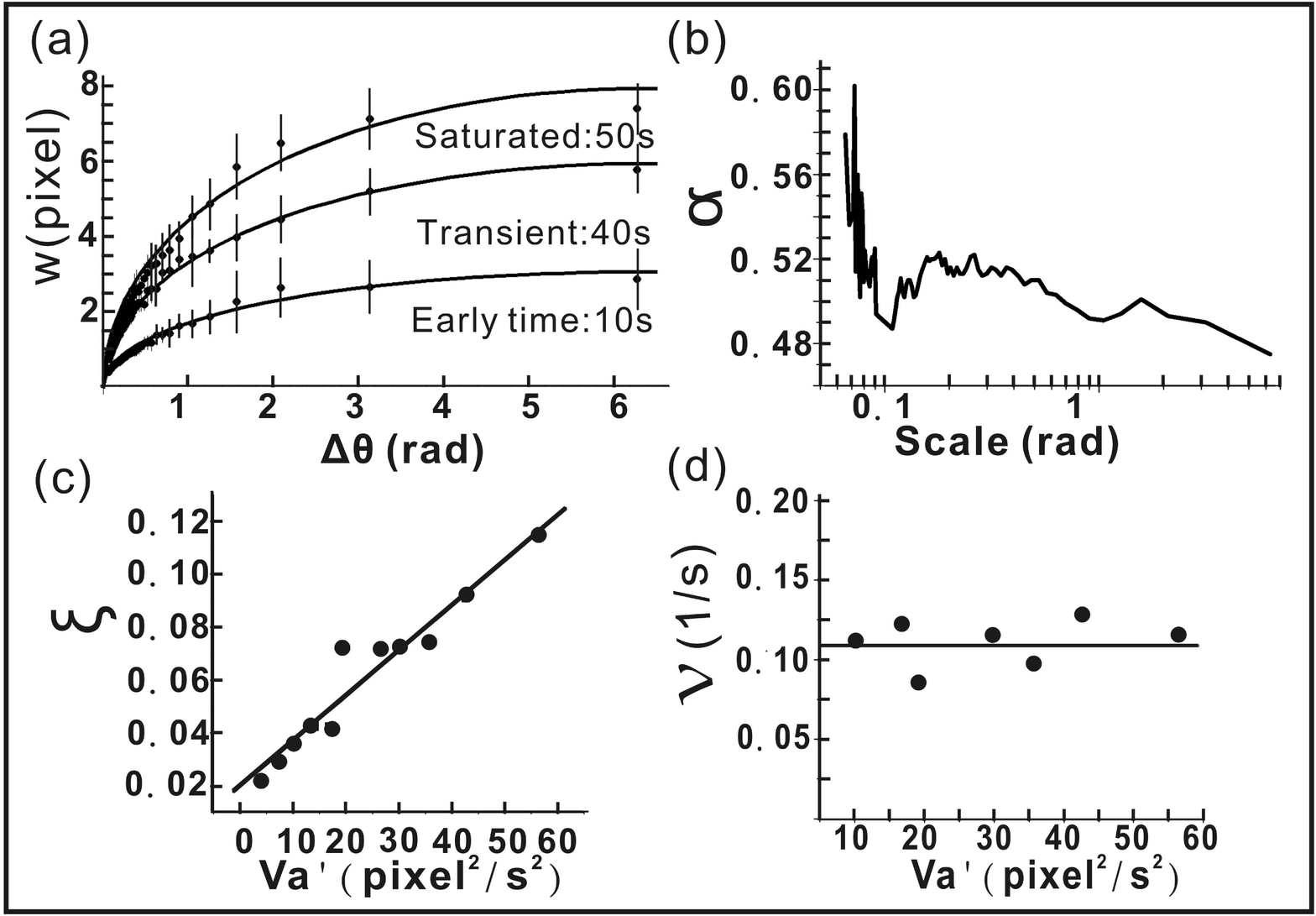}
\caption{\label{fig:combine2}(a) Scale dependence of a constant advancing interface with $V_{a}^{\prime}=4.01 \rm{ pixel^{2}/s^{2}}$. Solid line is the theoretical prediction. Because the flow rate is slow initially, more time is required for the designated early stage than in Fig.4. 
In (b), $\alpha$ is shown to vary with the scale.
 (c) $\xi$ is plotted against $V_{a}^{\prime}$. The slope of the fitting line, $6840$, falls within our predicted range. (d) Experimental values of $\nu$ are independent of $V_{a}^{\prime}$, consistent with our theory.}
\end{figure}

Note that with our previous choice of a constant injection speed the advancing interface is decelerating. In order to check that our conclusions are not its artifacts, we repeated by accelerating the injection flux to  attain a constant interface velocity. Now,   $\xi=\sigma \kappa V_{a}^{\prime}/\eta r^{2}$ becomes time-independent, $V_{a}$ is replaced by $V_{a}^{\prime}t$, but parameter $\nu$ is the same.
The experimental results in Fig.6-(a),(b) turn out to be qualitatively the same as Fig.3.
We find that $\xi$ is still proportional to $V_{a}^{\prime}$ with a slope that falls rightly within our estimated range $\sigma\kappa/\eta r^{2}=10^{3}\sim10^{4}$ (Fig.6-(c)) and $\nu$ remains independent of $V_{a}^{\prime}$ with the right magnitude (Fig.6-(d)).
We believe that if the value of $\kappa$ can be more accurately measured, the correctness of $\sigma\kappa/\eta r^{2}$ shall be more solidly verified.
We can now safely say that our dimensional analysis in the constant flow rate remains valid for the accelerated flow rate.

In conclusion, we pointed out that the roughness of a circular interface is intrinsically different from the standard planar injection. Major difference lies in the fact that the former interface grows in size.
This results in a noise spectrum whose amplitudes decrease with momentum, in contrast to the white noise for a planar interface. One major consequence is the lack of power law in the scaling and dynamic behavior, which we believe should be common to all circular patterns \cite{bacteria}. Interestingly, we also find that the time exponent varies with the range of scale, as has been reported \cite{scale5} for a planar interface but not captured by existing models. All these features can be described by a linearized model we modified from the EW equation. Two phenomenological coefficients are put forward to represent the competing trend of instability and smoothness. Their dependence on the microscopic parameters are determined from dimensional arguments, and estimated values conform with the experiments.

We benefit from discussions with Profs. H. H. Lin and C. K. Chan, and thank Prof. W. G. Wu for lending us his Isco 65D High Pressure Syringe Pump.
Support from NSC grants 95-2112-M007-046-MY3, 95-2120-M007-008, and 96-2815-C007-006 is acknowledged. 



\end{document}